\documentclass[aps, onecolumn, preprint, secnumarabic,nobalancelastpage,nofootinbib]{revtex4-1} 

\usepackage{graphics}      
\usepackage{graphicx}      
\usepackage{url}           
\usepackage{bm,color}            
\usepackage{amssymb, amsmath, enumerate, theorem, epsfig,subfigure}
\usepackage{physics}
\usepackage[normalem]{ulem}
\usepackage[table]{xcolor}
\usepackage{makecell}
\usepackage{verbatim}
\usepackage{upgreek}

\begin{document}

\title{All-optical cascadable universal logic gate with sub-picosecond operation}

\author{Anton V. Baranikov$^{1,\dagger}$, Anton V. Zasedatelev$^{1,\dagger}$, Darius Urbonas$^{2}$, Fabio Scafirimuto$^{2}$, Ullrich Scherf$^{3}$, Thilo St\"oferle$^{2}$, Rainer F. Mahrt$^{2}$, and Pavlos G. Lagoudakis$^{1,4}$ \footnote{Correspondence address:  \\ $\dagger$ A.V.B. and A.V.Z. contributed equally to this work}}
\date{\today}
\affiliation{$^1$Skolkovo Institute of Science and Technology, Moscow, Russian Federation }
\affiliation{$^2$IBM Research Europe - Zurich, S\"aumerstrasse 4, R\"uschlikon 8803, Switzerland }
\affiliation{$^3$Macromolecular Chemistry Group and Institute for Polymer Technology, Bergische Universit\"at Wuppertal, Gauss-Strasse 20, 42119 Wuppertal, Germany}
\affiliation{$^4$Department of Physics and  Astronomy, University of Southampton, Southampton, SO17 1BJ, United Kingdom}

\maketitle

\textbf{Today, almost all information processing is performed using electronic logic circuits operating with up to several gigahertz frequency. All-optical logic, however, that holds the promise to allow up to three orders of magnitude higher speed  \cite{Miller-NatPhot-2010} has not been able to provide a viable alternative because approaches that had been tried were either not scalable, not energy efficient or did not show a significant speed benefit. Whereas essential all-optical transistor functionalities have been already demonstrated across a range of platforms \cite{Chen-Science-2013, Ballarini-NatComm-2013, Hwang-Nature-2009}, using them to implement the complete Boolean logic gate set and in particular negation - switching off an optical signal with another optical signal - poses a major challenge \cite{Sun-NatPhoton-2019}. Here, we realize a universal NOR logic gate by introducing the concept of non-ground-state dynamic exciton-polariton condensation in an organic semiconductor microcavity under non-resonant pulsed excitation. In the presence of either of the input signals inserted at opposite in-plane momenta, non-ground state dynamic condensation supersedes spontaneous ground-state condensation, resulting in a NOR-operator output signal within less than a picosecond. An additional optical transistor, fed by the output of the NOR gate, is then used to regenerate the output signal such that it is usable as input for cascading gates, a prerequisite for scale up. Our results constitute an essential step towards the realization of more complex logic optical circuitry that could boost future information processing applications.
}

While the number of transistors that make up a processor has been growing exponentially over the last five decades the clock speed stalled at a few gigahertz about 15 years ago \cite{Waldrop-Nature-2016} as a result of the breakdown of Dennard scaling \cite{Fiori-NatureNanotech-2014}. The key challenge is that in order to keep the power density constant the dynamic dissipation must be further reduced when the frequency is increased. Current state-of-the-art transistors, although scaled down to single nanometer dimensions, typically require several attojoule switching energy. More energy efficient electronic devices, like single electron transistors have been investigated but were found to be incompatible with high speed, room temperature operation and established processing methods. On the other hand, optical devices like single photon all-optical transistors which offers only sub-attojoule switching energies have been realised \cite{Volz-NaturePhoton-2012,Sun-Science-2018}. But as they are based on epitaxially-grown quantum dots in photonic crystals at liquid helium temperature, they face similar roadblocks as their electronic counterparts, in addition to unsolved paths for cascadability and scale-up. Apart from the potential for reducing energy consumption whilst increasing switching speeds,  optical circuits could also harness a more precise clocking compared to electronics. Whereas density scaling of optical circuits is limited and probably will never reach the device densities of modern integrated electronics, already much simpler circuits compared to full-fledged microprocessors that could run at orders of magnitude higher speed while maintaining low power dissipation would be revolutionary, e.g. for data processing tasks that are today relying on application-specific integrated circuits (ASICs) \cite{Miller-JLightTechnol-2017}. 

The basic architecture of all-optical transistors usually involves an optical resonator --cavity, photonic crystal or other-- and a nonlinear optical process in the form of a higher order susceptibility, or an optical resonance in atomic gases or semiconductors. As such, semiconductor microcavities offer an apparent all-optical transistor platform using semiconductor excitons strongly coupled to a cavity mode leading to the formation of new hybrid light-matter eigenstates, denoted exciton-polaritons \cite{Weisbuch-PRL-1992, Kasprzak-Nature-2006}. Exciton-polariton transistor operation, including switching \cite{Ballarini-NatComm-2013, Gao-PRB-2012, Sturm-NatComm-2014}, amplification \cite{Saba-Nat-2001,Savvidis-PRL-2000}, and some logic gate functionality \cite{Ballarini-NatComm-2013,Marsault-APL-2015} were demonstrated in III-V semiconductor microcavities, alas at cryogenic temperatures. Most recently, utilising the properties of organic semiconductor polymers, room temperature all-optical exciton-polariton transistor operation was demonstrated with sub-picosecond switching time, a record net gain of $\sim$10 dB $\mathrm{\upmu{} m^{-1}}$, and cascadable all-optical AND/OR logic gate functionality \cite{Zasedatelev-NatPhot-2019}. However, the basic function of negation, an essential ingredient for complete logic has still remained elusive \cite{Sun-NatPhoton-2019}. Exciton-polariton transistors exploit the signal amplification that occurs through non-equilibrium exciton-polariton condensation, the dynamic formation of a macroscopically coherent wavefunction at the ground exciton-polariton state \cite{Deng-Science-2002, Daskalakis-nmat-2014, Cookson-AdvOptMat-2017}. The NOT gate would require coherent control of the exciton-polariton wavefunction \cite{Kundermann-PRL-2003}, a mechanism that necessitates stabilised interferometry and critical biasing on the level of the optical pulses involved, thus rendering such operation impractical for logic circuitry. 

Here, we demonstrate a universal exciton-polariton NOR logic gate by introducing the concept of non-ground-state dynamic exciton-polariton condensation in a configuration that does not rely on coherent interference of light beams. Furthermore, we show regeneration of the output signal to make it usable as input for cascaded multiple gates. Such element allows to construct any combinatorial logic function and also larger, more complex logic, a key ingredient for future all-optical logic circuitry.

To scrutinize the formation of non-ground-state dynamic exciton-polariton condensation, we employ a semiconductor polymer microcavity consisting of a 35 nm thick neat film of methyl-substituted ladder-type poly-[paraphenylene] (MeLPPP) embedded between 50 nm SiO$_{2}$ spacers, sandwiched between SiO$_{2}$/Ta$_{2}$O$_{5}$ distributed Bragg reflectors (DBR) deposited on an optically flat fused silica substrate. Strong coupling of the cavity mode (2.65 eV) and two sub-levels of the first excited singlet state (S$_{10}$ at 2.72 eV and S$_{11}$ at 2.91 eV) result in three exciton-polariton branches, shown in Fig.1a, exhibiting 144 meV Rabi splitting between the middle and the lower exciton-polariton branches \cite{Plumhof-NatureMat-2014}. Under non-resonant optical pumping at condensation threshold the full-width at half-maximum (FWHM) of the ground-state exciton-polariton wavefunction in Fourier-space is $\sim$0.49$\mathrm{\upmu{}m^{-1}}$ (see Fig.S1b). By configuring a seed pulse to resonantly inject exciton-polaritons with non-zero in-plane momenta we stimulate non-ground state dynamic exciton-polariton condensation. Furthermore, we minimize the angular overlap between the ground-state exciton-polariton condensate (corresponding to the gate output signal) and the seed pulse (corresponding to the gate input signal). This is done, while retaining their energy difference, by tuning the energy of the seed in resonance with the lower exciton-polariton branch at an in-plane wavevector of $\sim$-2.55$\mathrm{\upmu{}m^{-1}}$, see Fig.1a and Methods. In addition, we tune the energy of the non-resonant optical pump one molecular vibron energy above the energy of the seed pulse to enable a single-step vibron-mediated energy relaxation from the hot exciton reservoir, as shown in Fig.1a. The non-resonant pump excitation density dependence of the emission intensity resonant to the seed pulse at zero pump-seed time delay is shown in Fig.1b. The horizontal dashed line indicates the transmission intensity of the seed-only and the grey shaded area the threshold excitation density, see Methods. Figure 1c displays the corresponding FWHM and the emission energy at the maximum of the emission spectrum. At threshold, we observe a collapse of the FWHM and an energy blue-shift, indicative of the formation of an exciton-polariton condensate resonant to the seed. A dispersion image of the emission above pump threshold is shown in Fig.1a, overlaid with the calculated exciton-polariton branches.

To identify the influence of the seed beam we examine the emission intensity dependence at the ground exciton-polariton state on the seed power. In the absence of the seed pulse and at an excitation density approximately twice above threshold for dynamic condensation $P_{th}$, the non-resonant pump induces ground-state exciton-polariton condensation, see Supplementary Information (SI), Section I. Figure 2a shows the emission intensity of an exciton-polariton condensate at the ground state angularly filtered at normal incidence, \textit{k}$_{\parallel}$ = 0, with a width of  $\sim$2$\mathrm{\upmu{}m^{-1}}$ versus seed-power for zero pump-seed time delay. The pump excitation density is kept constant at $P\sim$2$P_{th}$ for unseeded condensation driving the system in the saturation regime to minimize noise induced by laser intensity fluctuations. With increasing the seed power, a clear threshold, depicted with a grey shaded area, above which the emission intensity decreases monotonically is observed. In the presence of the seed pulse, exciton-polariton relaxation to the non-ground polariton state is more efficient than to the ground exciton-polariton state, resulting in an extinction coefficient of $\sim$50$\times$, calculated by the ratio of the integrated emission intensity with and without the seed pulse. This dependence illustrates NOT gate functionality, where the pump beam charges the transistor and the seed beam acts as the control input that switches the output between ``1" and ``0" states.

Building on the concept of non-ground-state dynamic exciton-polariton condensation, we demonstrate a universal NOR gate. We configure a two-input NOR gate by adding a second seed beam injecting exciton-polaritons resonantly with opposite in-plane momentum, i.e. \textit{k}$_{B}$ = -\textit{k}$_{A}$, as shown schematically in Fig.2b keeping the power of the two seed beams equal across the whole study, see Methods. Figure 2b shows the emission intensity of the exciton-polariton condensate at the ground state angularly filtered at normal incidence, \textit{k}$_{\parallel}$ = 0, with a FWHM of  $\sim$2$\mathrm{\upmu{}m^{-1}}$ versus seed-power for zero pump-seeds time delay. Similarly, to the single seed experiment, with increasing the seeds' power we observed a clear threshold, depicted with a grey shaded area, above which the emission intensity decreases monotonically. This two seed beam configuration exhibits an extinction coefficient of $\sim$60$\times$. Evidently the two seed beams act as control inputs to the NOR gate. Figure 3 shows the spatial profile of the emission intensity for the four input configurations of the two seed beams, and the corresponding dispersions demonstrating that in the presence of either of the inputs, non-ground-state dynamic condensation supersedes spontaneous ground-state condensation resulting at the truth-table of the universal NOR gate.

In Figure 3, we note a difference between the emission energy of ground-state and non-ground-state exciton-polariton condensates; 2,595 eV and 2,606 eV respectively. Such mismatch would now allow for cascading successive NOR gates. To circumvent this limitation, we restore the output of the NOR gate, shown in Fig.4a by cascading and re-conditioning the output of the NOR gate via adding another transistor stage realized by a non-resonant optical pump identical to that of the gate (for details see SI, Section II). Figure 4b shows the emission spectrum of the output from the NOR gate whereas in Fig.4c the emission of both the re-conditioned ground-state exciton-polariton condensate (in red) and of the non-ground-state condensate (in green) is shown. Apparently, the process of cascaded amplification induces an additional energy-shift to the emission of the output signal \cite{Yagafarov-CommPhys-2020}, allowing to match the one of the input signals' photon energy. Such NOR gate with regenerated output exhibits all inputs and outputs at the same photon energy and provides enough power to fan-out to multiple subsequent gates, a prerequisite for scaling up to complex circuits.

An important advantage over previous realizations of exciton-polariton AND/OR gates \cite{Zasedatelev-NatPhot-2019} is that the repetition rate of this universal NOR gate is not thwarted by the relatively slow dynamics when no condensate state is reached, i.e. for a logic ``0'' output. With the present NOR gate, an exciton-polariton condensate is always in place independently of the logic levels, but formed at distinctly different wavevectors for ``0'' and ``1'' through rapid depopulation of the excited state, i.e. with sub-picosecond response time, $\sim$500 fs (for details see SI, Section III). Noteworthy to mention that it is straightforward to extend the method to multi-input joint denial truth-functional operators utilising the full two-dimensional parabolic exciton-polariton dispersion.

The realisation of a universal exciton-polariton NOR gate provides the basic building block for a complete all-optical logic circuitry platform, operational at room temperature and high speed. Future scaling up to complex logic circuits will favourably harness planar architectures of photonic microcavities that are readily available.

\noindent


\clearpage

\begin{figure}[t]
	\centering
	\includegraphics[width=18.3cm]{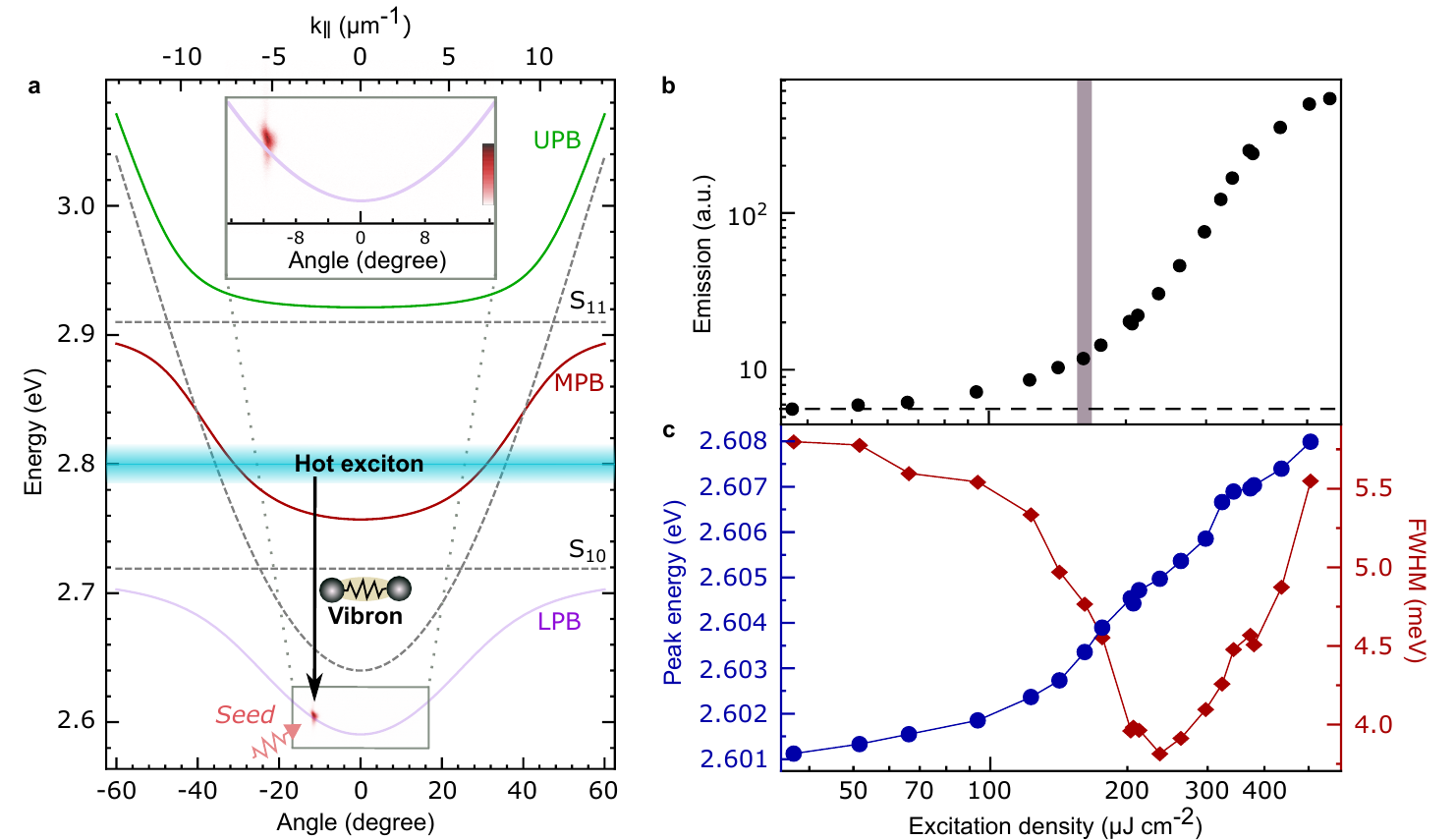}
	
	\label{1}
\end{figure}

\clearpage

\begin{figure}[t]
	\centering
	\includegraphics[width=8.9cm]{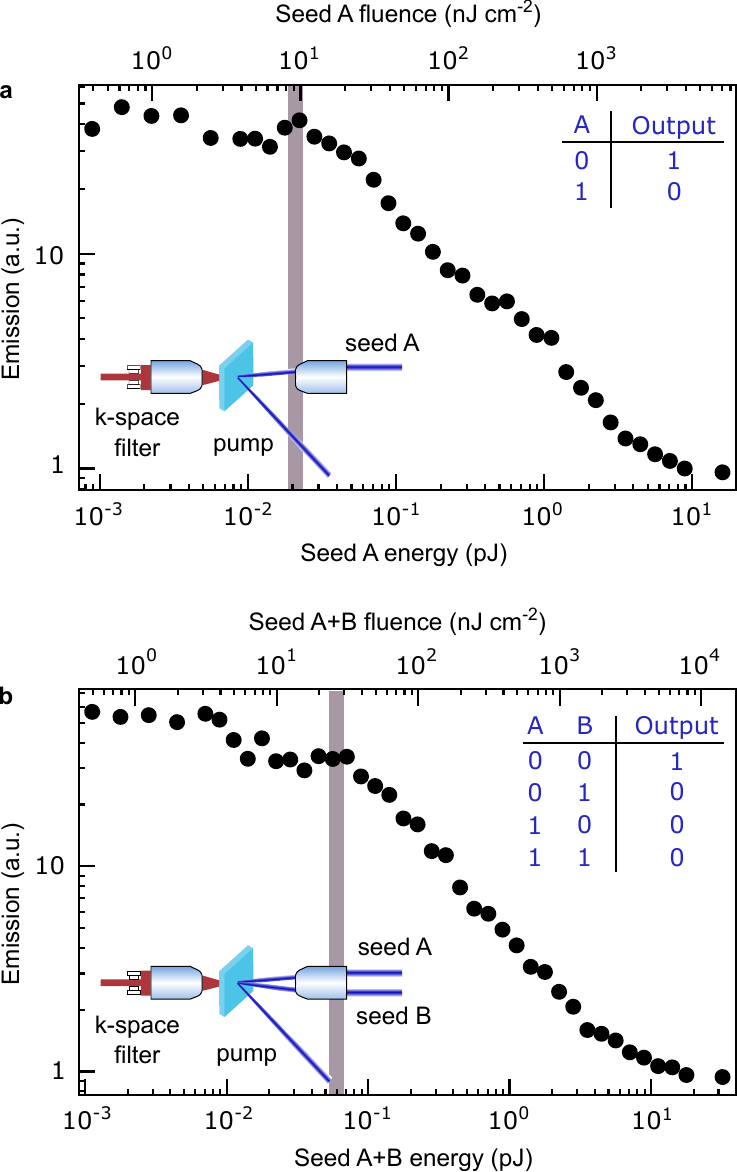}

	\label{2}
\end{figure}

\clearpage

\begin{figure}[t]
	\centering
	\includegraphics[width=18.3cm]{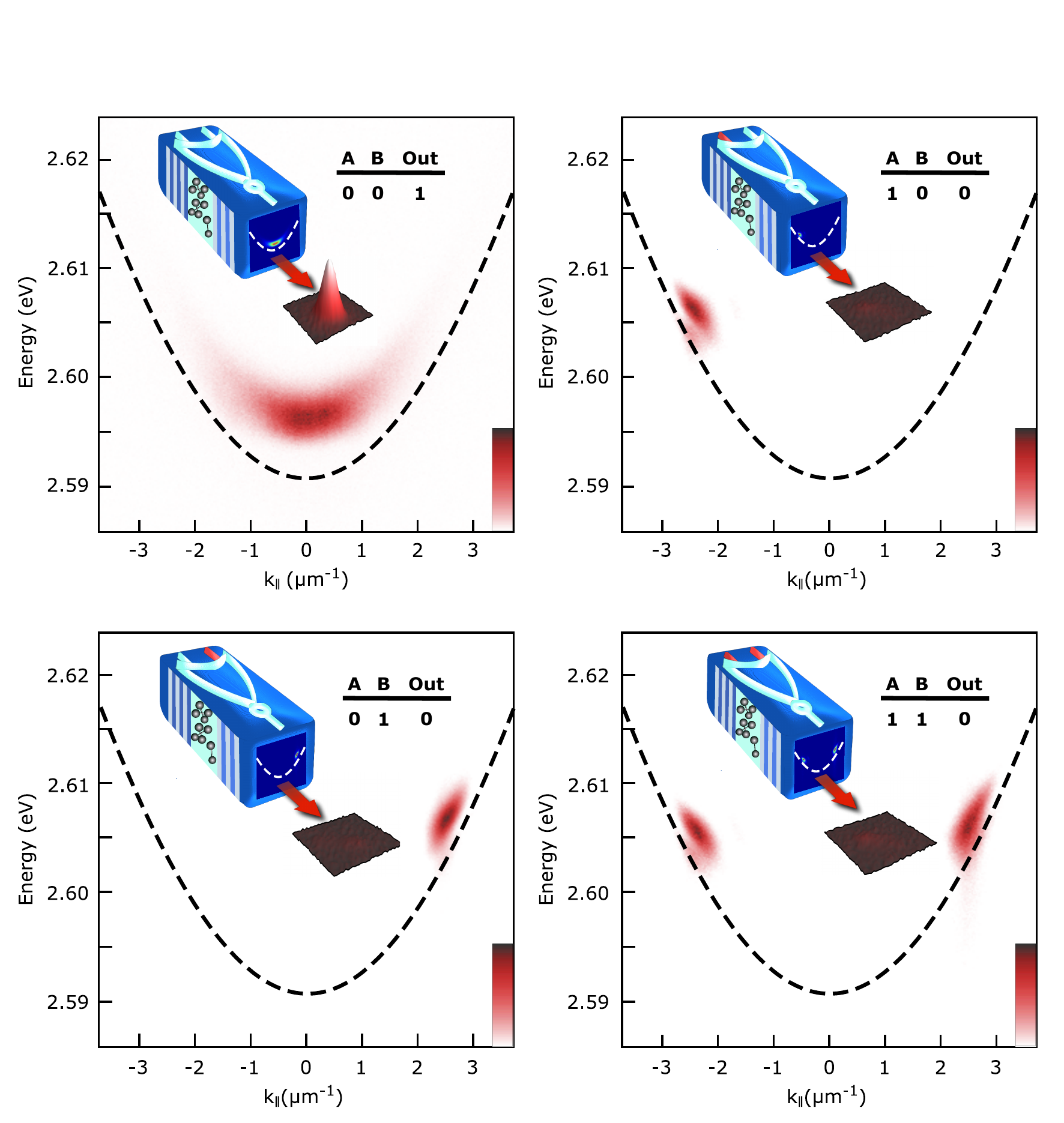}
	
	\label{3}
\end{figure}

\clearpage

\begin{figure}[t]
	\centering
	\includegraphics[width=8.9cm]{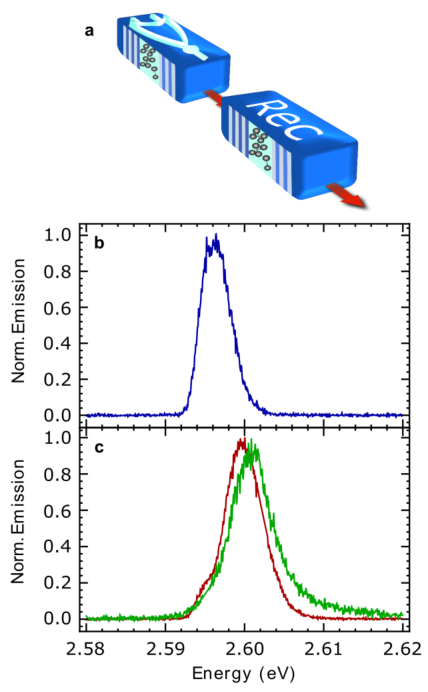}
	
	\label{4}
\end{figure}

\clearpage

\section*{Figures captions}

\textbf{Fig. 1: Non-ground-state dynamic exciton-polariton condensation.} \textbf{a}, Strong coupling of the optical cavity mode and the two sub-levels of the first excited singlet state of the organic semiconductor (horizontal dashed lines, S$_{10}$ and S$_{11}$) result in three exciton-polariton branches - the upper/middle/lower branch drawn in green/red/violet color respectively versus in-plane wavevector (polariton in-plane momentum-- top axis) and angle (with respect to the normal to the plane of the cavity-- bottom axis). The optical pump is tuned at 2.8 eV injecting ``hot" excitons depicted with a blue-shaded area. A seed beam injects few exciton-polaritons resonantly at \textit{k}$_{\parallel}$ = $\sim$-2.55$\mathrm{\upmu{}m^{-1}}$. The exciton-cavity detuning is chosen to align the seeded exciton-polariton state at one vibronic energy quantum of 200 meV below the pump-injected hot-excitons. Energy relaxation from the hot-exciton reservoir occurs with the emission of a single vibron (black solid vertical arrow) and is stimulated by the incident optical seed depicted with a red-arrow. Superimposed with the calculated lower exciton-polariton dispersion within the dotted frame is the normalised emission intensity in the pump-seed configuration at $P_{pump}$ = 200 $\mathrm{\upmu{}}$J cm$^{-2}$. \textbf{b}, Emission from \textit{k}$_{\parallel}$ = $\sim$-2.55$\mathrm{\upmu{}m^{-1}}$ angularly integrated over $\sim$ \textit{k}$_{\parallel}\ni$ (-2.96;-1.86) $\mathrm{\upmu{}m^{-1}}$ versus pump excitation fluence at zero pump-seed time delay. The grey shaded area indicates the dynamic condensation threshold and the horizontal dashed line the level of the recorded intensity from the seed-only. \textbf{c}, Emission photon energy at the maximum of the emission spectrum (left/blue axis) and full-width at half-maximum (FWHM, right/red axis) versus pump excitation fluence at zero pump-seed time delay. 	
	
\textbf{Fig. 2: Depletion of the ground exciton-polariton state.} \textbf{a},\textbf{b}, Ground-state condensate emission filtered over $\sim\pm$1$\mathrm{\upmu{}m^{-1}}$ versus seed energy and seed fluence at zero pump-seed time delay for one \textbf{a} and two \textbf{b} seed pulses. The seed pulses are of equal power and inject exciton-polaritons resonantly at opposite in-plane wavevectors (\textit{k}$_{\parallel}$ = $\sim\pm$2.55$\mathrm{\upmu{}m^{-1}}$). The grey shaded areas indicate seed threshold power. The inset tables in \textbf{a} and \textbf{b} are the truth tables for the NOT and NOR gates respectively.

\textbf{Fig. 3: Universal exciton-polariton gate.} Normalized emission of the nonlinear lower exciton-polariton branch in the four control input configurations annotated in each panel. The black dashed line is a parabolic fit to the linear dispersion, illustrating the blue-shift of the exciton-polariton dispersion in the nonlinear regime. At the facet of each schematic we overlay the corresponding exciton-polariton dispersion image and the red arrows point to the corresponding spatial emission profile ($\sim$20x20 micrometers).
	
\textbf{Fig. 4: Output signal regeneration.} \textbf{a}, The output of the NOR gate is reconditioned through cascaded re-amplification of the ground-state emission of an exciton-polariton condensate in the presence of a second non-resonant optical pump identical to that of the NOR gate. \textbf{b}, The emission spectrum of the output of the NOR gate without regeneration. \textbf{c}, The emission spectrum of both the re-amplified ground-state (in red) and of the non-ground-state exciton-polariton condensates (in green), showing the good match between the regenerated gate output spectrum and the input spectrum.

\clearpage

\section*{Methods}

\textbf{Sample fabrication}.

The sample is composed of a bottom distributed Bragg reflector (DBR), a central cavity defect region with an effective thickness slightly larger than half the exciton wavelength, and a top DBR on a fused silica substrate. The DBRs consist of alternating SiO$_{2}$/Ta$_{2}$O$_{5}$ quarter-wavelength-thick layers produced by sputter deposition (9+0.5 pairs for the bottom DBR, 6+0.5 for the top DBR). The center of the cavity consists of a polymer layer sandwiched within 50-nm spacer layers of sputtered SiO$_{2}$. The spacer is deposited on the organic using a SiO$_{2}$ sputter target. Methyl-substituted ladder-type poly(p-phenylene) (MeLPPP; $M_{n}=31500$, $M_{w}=79000$) was synthesized as described elsewhere \cite{Scherf-Macromolecularchemistry-1992}. MeLPPP is dissolved in toluene and spin-coated on the bottom spacer layer. The film thickness of approximately 35 nm is measured with a profilometer (Veeco Dektak).

\textbf{Spectroscopy}.
The pump pulse with $\sim$200 fs duration is provided by a tunable optical parametric amplifier (Coherent OPerA SOLO) excited by 500 Hz high energy Ti:Sapphire regenerative amplifier (Coherent Libra-HE). The beam is spectrally filtered, providing 25 meV full-width at half-maximum (FWHM). The pulses are focused on the sample with a 50 mm lens at oblique incidence (45$^{\circ}$). The pump has an elliptical profile with $\sim$40 $\mathrm{\upmu{}m}$ and 26 $\mathrm{\upmu{}m}$ spot sizes. The seed beams are produced by generating white light continuum (WLC) in a sapphire plate excited with 800 nm ultrashort ($\sim$100 fs) pulses. WLC is then spectrally filtered, resulting in broadband (2.59-2.7 eV) emission. The seed pulses are focused on the sample by an objective (10x Nikon, 0.3 NA), resulting in a spot size of $\sim$25 $\mathrm{\upmu{}m}$. A motorized translation stage with a retroreflector allows to adjust the temporal delay between the pump and seed pulses. For the complete experimental setup see SI, Section II.

Momentum- and energy-resolved emission is acquired in the transmission configuration. Output emission of the sample is collected with an objective (10x Mitutoyo Plan Apo, NA = 0.28) and coupled to a spectrometer (Princeton Instruments SP2750) with an electron multiplying charge coupled device (EMCCD) camera (Princeton Instruments ProEM 1024BX3). The emission was spectrally resolved using a 1200 grooves/mm grating and a slit width of 200 $\mathrm{\upmu{}m}$ at the entrance of the spectrometer. An additional 1000 mm conjugated lens is used to project the Fourier plane of the collecting objective to the slit. The pump power dependence measurements (Fig.1b,c) are obtained through integrating the output emission within $\sim$ \textit{k}$_{\parallel}\ni$ (-2.96;-1.86) $\mathrm{\upmu{}m^{-1}}$ around \textit{k}$_{probe}$. The fluence of the seed beam is fixed at 45 nJ cm$^{-2}$. To obtain the incident excitation density of the pump pulse, the average pump power is measured using a calibrated Si photodetector (Thorlabs-Det10/M) and an oscilloscope (Keysight DSOX3054T). Accuracy verification of the power measurements is carried out by using a powermeter: Si photodiode power sensor (Thorlabs-S120VC) with a console (Thorlabs-PM100D).

The seed power dependences (Fig.2) are taken under pumping at $P\sim$2$P_{th}$ for unseeded exciton-polariton condensation. Both seed beams have an incident angle of $\sim$11$^{\circ}$ with the opposite momenta, e.g. \textit{k}$_{B}$ = -\textit{k}$_{A}$ $\sim$-2.55$\mathrm{\upmu{}m^{-1}}$. The real space data depicting the NOR gate functionality (Fig.3) is obtained with angular filtering of the Fourier plane using an iris aperture.

\clearpage

\section*{Data availability}
All data supporting this study are openly available from the University of Southampton repository at https://doi.org/10.5258/SOTON/D1142.

\section*{Acknowledgements}
This work was partly supported by QuantERA project RouTe, the Swiss State Secretariat for Education, Research and Innovation (SERI) and the European Union Horizon-2020 framework program through the Marie-Sklodowska Curie ITN network SYNCHRONICS (H2020-MSCA-ITN-643238) and the UK’s Engineering and Physical Sciences Research Council grant EP/M025330/1 to P.G.L on Hybrid Polaritonics and the Russian Scientific Foundation (RSF) grant No. 18-72-00227 to A.Z. and the RFBR according to the research projects No. 20-52-12026 (jointly with DFG) and No. 20-02-00919.

\section*{Author contributions}
A.V.Z. and A.V.B. performed the experiments and analysed the experimental data. D.U., F.S., T.S., and R.F.M. contributed to the design and fabrication of the organic microcavity. U.S. synthesised the organic material. P.G.L. designed and led the research. All authors contributed to the writing of the manuscript and have given their approval to the final version of the manuscript.

\section*{Additional information}
The authors declare no competing financial interests.

\clearpage

\newcommand{\nb}[1]{\textcolor{black}{#1}}
\newcommand{\rrr}[1]{\textcolor{black}{#1}}
\newcommand{\ppp}[1]{\textcolor{black}{#1}}
\newcommand{\qqq}[1]{\textcolor{black}{#1}}
\newcommand{\zzz}[1]{\textcolor{red}{#1}}

\setlength{\arrayrulewidth}{1mm}
\setlength{\tabcolsep}{13pt}
\renewcommand{\arraystretch}{1.5} 
\newcolumntype{P}[1]{>{\centering\arraybackslash}m{#1}}
 \setcounter{figure}{0}
\renewcommand{\thefigure}{S\arabic{figure}}
 \setcounter{table}{0}
\renewcommand{\thetable}{S\arabic{table}}

\textbf{\large{Supplementary Information: All-optical cascadable universal logic gate with sub-picosecond operation}}



\section{Unseeded exciton-polariton condensation}
Intense vibronic resonances of MeLPPP render exciton-polariton condensation easy even without any prearrangement in exciton-polariton occupancy. Relying on the single-step vibron-mediated exciton-to-polariton relaxation process, we realise unseeded exciton-polariton condensation that we harness in our binary logic as the high level - "1". Figure S1a shows a schematic of the optical excitation that forms a hot exciton reservoir, which is superimposed with the dispersion relation of the exciton-polariton system under investigation. Note the energy of hot excitons (2.8 eV) here is one vibronic energy quantum (0.2 eV) above the ground exciton-polariton state (2.6 eV). Emission energy-momentum distributions recorded below ($P\sim$100 $\mathrm{\upmu{}}$J cm$^{-2}$) and above ($P\sim$280 $\mathrm{\upmu{}}$J cm$^{-2}$) the threshold are shown in Fig.S1b as the top and bottom panels respectively. Analysis of the emission integrated over \textit{k}$_{\parallel}$ = 0 $\pm$ 1.6 $\mathrm{\upmu{}m^{-1}}$ demonstrates superlinear rise of exciton-polariton occupancy above the threshold of $\sim$230 $\mathrm{\upmu{}}$J cm$^{-2}$ (Fig.S1c top panel) accompanied with significant line narrowing and blueshift (Fig.S1c bottom panel) - commonly recognized features of exciton-polariton condensation in organic microcavities.

 \begin{figure}[h!]
	\centering
	\includegraphics[width=18.3cm]{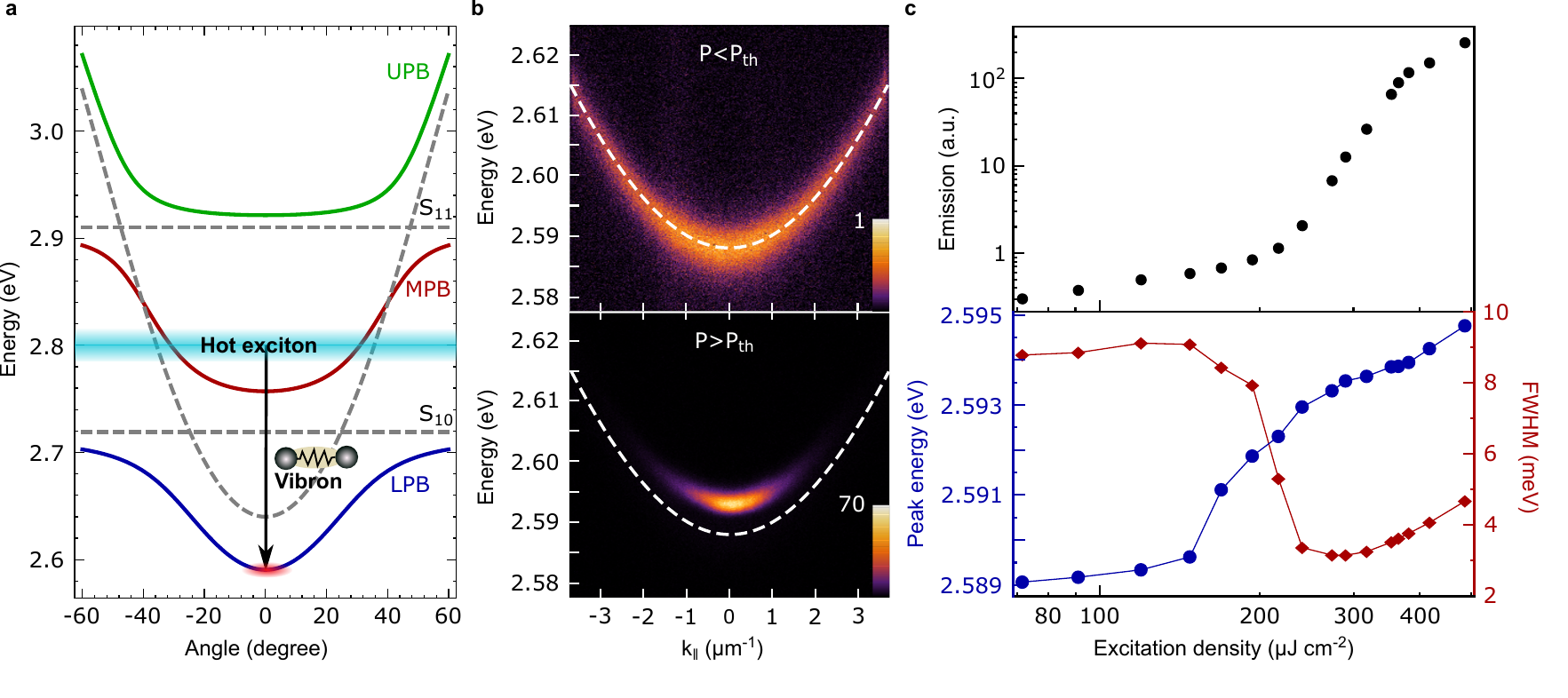}\\       	
	\caption{\textbf{Unseeded exciton-polariton condensation.} \textbf{a}, Schematic of the optical excitation (hot excitons) superimposed with upper/middle/lower exciton-polariton branches drawn in blue, red, and green solid lines, respectively. Dashed lines demonstrate the bare cavity mode (Cav) and the two sub-levels of the first excited singlet state of MeLPPP (S10 and S11). The optical pump is aligned at 45 degrees incidence with 2.8 eV photon energy effectively injecting ``hot" excitons depicted with a blue-shaded area. The vibron-mediated hot exciton-to-polariton relaxation process is depicted by a black solid vertical arrow. \textbf{b}, The top and the bottom panels show $E,k$-distribution of the emission below ($P\sim$100 $\mathrm{\upmu{}}$J cm$^{-2}$) and above ($P\sim$280 $\mathrm{\upmu{}}$J cm$^{-2}$) the condensation threshold respectively, where the dashed line corresponds to the unperturbed lower exciton-polariton branch.\textbf{c}, The top panel depicts the output emission integrated over \textit{k}$_{\parallel}$ = 0 $\pm$ 1.6 $\mathrm{\upmu{}m^{-1}}$ versus excitation density, while the bottom panel shows the spectral characteristics of the emission, namely its peak position (red) and the full-width half-maximum (FWHM, blue).
	}
	
\end{figure}

\clearpage

\section{The experimental setup}
All the experiments described in the main text (Fig.1 and Fig.2) have been carried out using the following experimental arrangement. 
To realise NOT and NOR gates operation at room temperature, we configure non-ground state dynamical exciton-polariton condensation. As seed beams, which serve is inputs for the gates, we employ a white-light continuum (WLC) generated in a sapphire plate split into two beams with equal intensity, the seed beam "A" and "B". Both the seed beams are focused on the sample by the same objective. To access non-ground exciton-polariton states within the lower branch, we seed the system under a different angle of incidence that we control through the translation stages in the optical paths of both seed beams accordingly. To control the temporal overlap between the pump and both seed beams, we utilise two motorized delay lines - translation stages equipped with retroreflectors, as shown in Fig.S2. The output emission is filtered in the Fourier plane of the objective by a home-made short pass \textit{k}$_{\parallel}$-filter, which cuts the light with in-plane momentum $\vert$\textit{k}$_{\parallel}\vert>$ 1 $\mathrm{\upmu{}m^{-1}}$. Note that the real space distributions in Fig.3 represent the intensity of emission collected at \textit{k}$_{\parallel}$=0 reflecting the population of the ground exciton-polariton state. We record this data using an iris aperture in the Fourier plane as a two-dimensional $kx, ky$ low pass filter.

We have filtered the WLC beam spectrally by a short pass filter to avoid seeding the lower-lying exciton-polariton states through possible leaky modes and sample defects (in particular states around \textit{k}$_{\parallel}$=0). Figure S3 shows the spectrum of the seed beams with respect to typical energy bands of the ground state as well as non-zero momentum exciton-polariton states under investigation depicting the complete suppression of parasitic scattering from the seed beams in the vicinity to the ground exciton-polariton state at \textit{k}$_{\parallel}$=0 (2.59 eV). 

We further extend the functionality of the exciton-polariton NOR gate by investigating the tolerance of the effect with respect to the angle of incidence for the seed beams. We measure the extinction for the NOR gate, i.e. the ratio between high and low level output, expressed in dB versus in-plane momentum of resonantly injected exciton-polaritons by one of the seed beams. Figure S4 shows the flexibility of the gate under the broad range of seeded states, which allows for tuning the energy of the seed beams. As expected, the extinction gradually decreases with increasing \textit{k}$_{\parallel}$ (the bottom x-axis) as the resonance with the vibron for higher energy exciton-polariton states is lost. 

The feasibility for signal photon energy restoration is an important extension of the exciton-polariton NOR gate functionality. We harness the ubiquitous high-energy shift phenomenon of exciton-polariton condensates, so-called blueshift \cite{Yagafarov-CommPhys-2020}, to adjust the gate output photon energy towards seed pulses control the gate. We implement the second amplification stage (B) which is pumped with the twice higher fluence of $P\sim$4$P_{th}$ as shown in Fig.S5a. The strong pump beam serves as the adjusting knob biasing photon energy of the output signal towards the seed beam and makes fine-tuning feasible within $>10$ meV energy range via substantial blueshift of the ground exciton-polariton state. Figure S5b shows emission spectra of the output signal before (A) and after (B) restoration process, one can observe significant amplification of the initial gate output that is byproduct of the photon energy tuning. The output of stage B can be used as the seed for next exciton-polariton NOR gates.

\begin{figure}[h!]
	\centering
	\includegraphics[width=8.9cm]{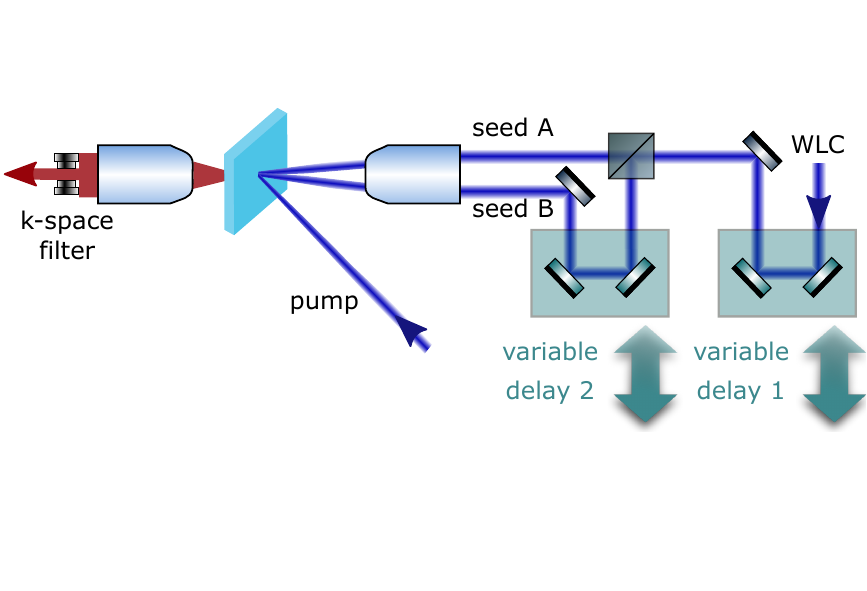}\\       	
	\caption{\textbf{Scheme of the experiment.} Filtered WLC is split into two beams, seed A and B. We use variable delay lines in each arm to maximize temporal overlap between pump and both seed beams.
	}
\end{figure}

\begin{figure}[h!]
	\centering
	\includegraphics[width=8.9cm]{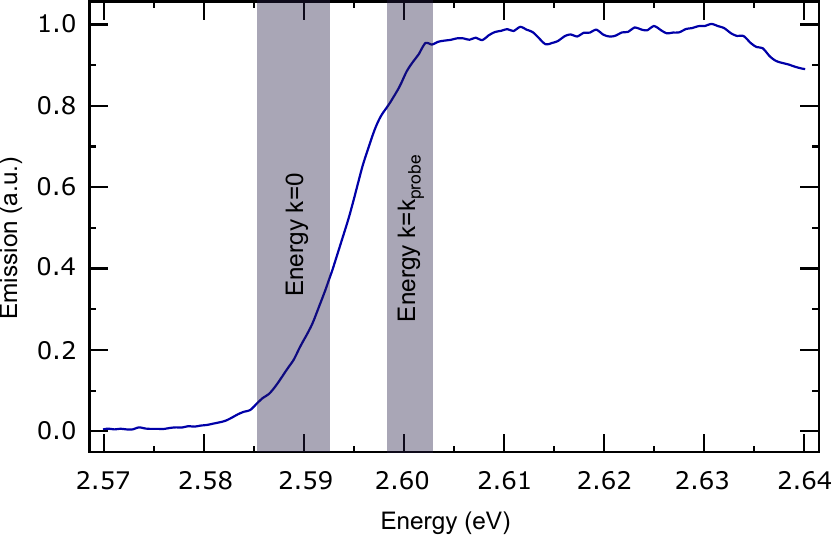}\\       	
	\caption{\textbf{The seed spectrum.} The spectrum of seed beams (blue) after passing short pass filters that cut out low energy photons to prevent for seed scattering at the energy corresponding to \textit{k}$_{\parallel}$=0. Gray-shaded areas show typical energy bands of the ground state as well as non-zero momentum exciton-polariton states under investigation.
	}
\end{figure}

\begin{figure}[h!]
	\centering
	\includegraphics[width=8.9cm]{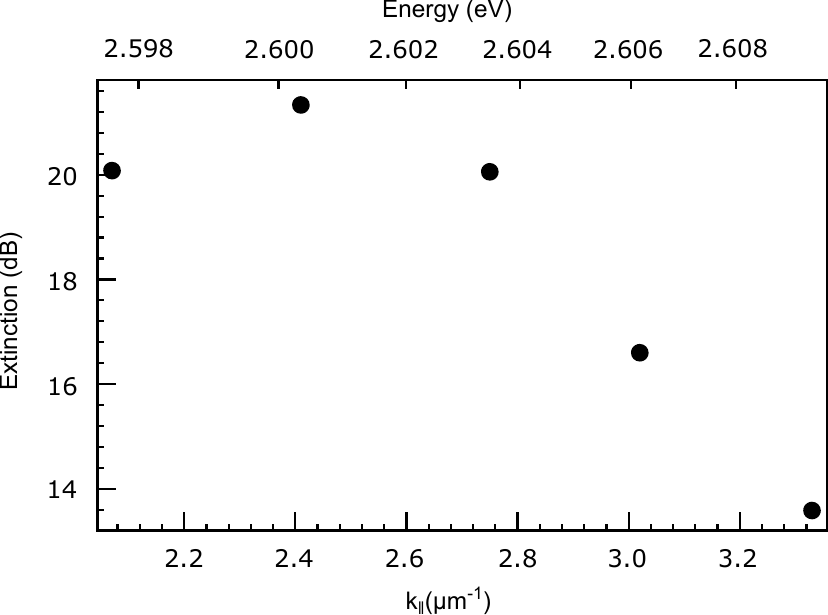}\\       	
	\caption{\textbf{Flexibility of the exciton-polariton NOR gate.} The gate operates at room-temperature within a broad range of seeded states, which allows for tuning the energy of seed beams. The extinction gradually decreases with increasing \textit{k}$_{\parallel}$ (the bottom x-axis) as one loses the resonance condition with the vibron at higher energy exciton-polariton states. The top x-axis shows the energy of the corresponding exciton-polaritons states resonantly injected by the seed beam.
	}
\end{figure}

\begin{figure}[h!]
	\centering
	\includegraphics[width=12.8cm]{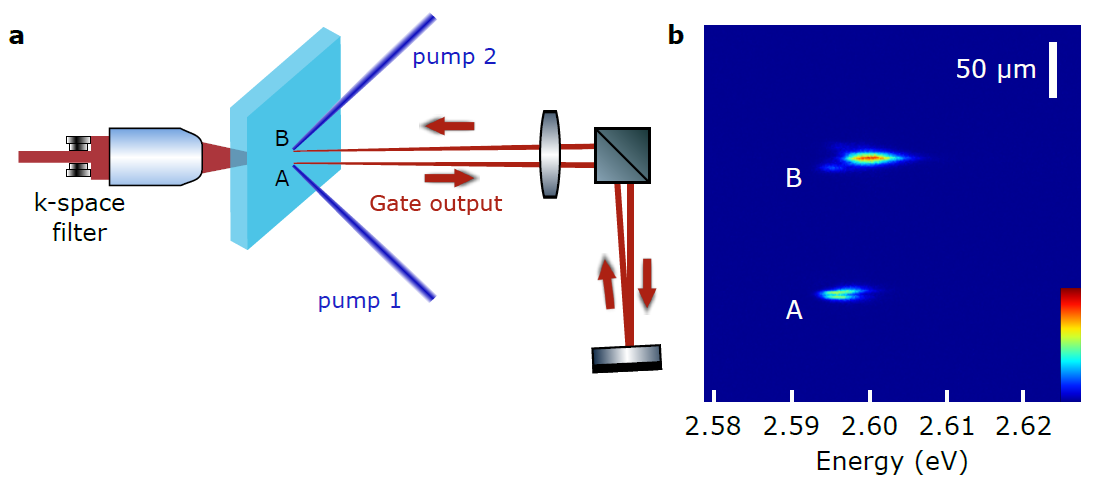}\\      
	\caption{\textbf{Scheme of signal restoration } \textbf{a}, Signal restoration of the gate output photon energy at stage A is feasible via blue-shifted emission at stage B within the re-amplification process. The gate output from the first stage A is coupled to the stage B by a free-space optical setup allowing for synchronization Pump 2 beam with the delayed gate output signal. \textbf{b}, Energy-resolved spatial distribution along vertical axis shows emission spectra at the stage A and B filtered in a Fourier plane of objective within $\vert$\textit{k}$_{\parallel}\vert<$ 1 $\mathrm{\upmu{}m^{-1}}$ in-plane momentum range.  
	}
\end{figure}

\clearpage

\section{Ultra-fast switching of the universal gate}

We investigate the transient dynamics of the gate operation by delaying the seed pulse against the pump pulse. Figure S5 shows emission measured at \textit{k}$_{\parallel}$=0 as a function of the delay. In contrast to the stimulated exciton-polariton condensation into the ground state reported in \cite{Zasedatelev-NatPhot-2019}, we observe the reverse effect, i.e. depletion of the exciton-polariton occupancy at the state in the vicinity of zero-time delay. The decrease in occupancy is the result of the competitive process, namely non-ground state stimulation triggered by the seed pulse. Quantitative comparison of the transient dynamics by means of the full-width half-maximum/minimum for the stimulated ground and non-ground state exciton-polariton condensation, respectively, indicates a similar response time of $\sim$500 fs. The ultra-fast exciton reservoir dynamics governed by the process of stimulated condensation as well as the short exciton-polariton lifetime $\sim$100 fs result in sub-picosecond switching time. Thus, the speed of the exciton-polariton gate is several orders of magnitude higher compared to current conventional optical logic platforms. We provide a comparative analysis in the next section.

\begin{figure}[h!]
	\centering
	\includegraphics[width=8.9cm]{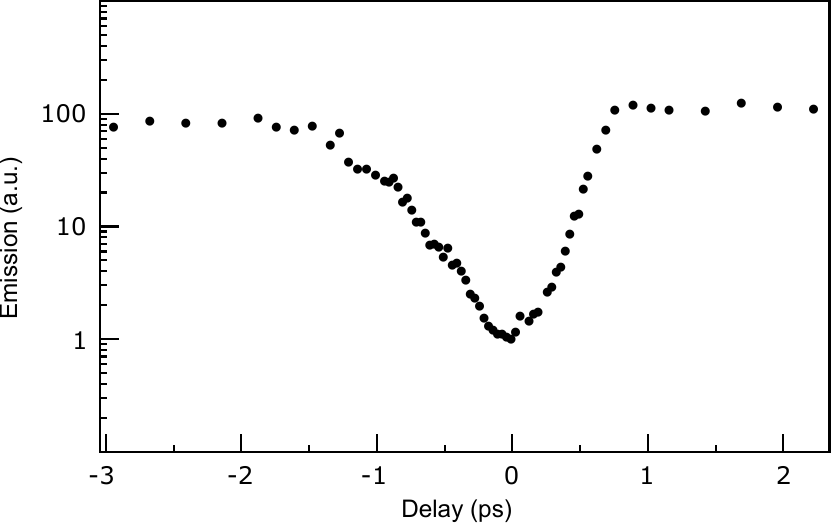}\\       	
	\caption{\textbf{Transient dynamics of the universal gate.} Emission versus time delay between the seed and pump pulses. The full width of half-minimum of the temporal response is $\sim$500 fs. The pulse energy of the control beam (i.e. the switching energy) is $\sim$10 pJ.
	}
\end{figure}

\clearpage

\section{NOR optical gates}

In this section, we provide a brief overview of the most advanced NOT/NOR optical gates. Several platforms have been developed for optical circuitry to date. Generally, all of them rely on nonlinear optical phenomena modulating an optical output by means of weak optical control signal input. We deliberately do not include any interferometric configurations. As being phase-sensitive, they impose additional stringent requirements discussed in the main text. It is reasonable to divide all-optical universal gates in two large general classes: 1 - based on semiconductor optical amplifiers (SOA), 2 - other platforms, including fiber and wave-guide technologies.

\textbf{Gates based on SOA.} SOA-based optical gates usually rely on cross-gain (XGM) or cross-phase (XPM) modulation. For the former principle, one needs to realise the multi-mode regime in which a particular mode amplification decreases with an increase of input signal intensity. This becomes possible once optical inputs are cross-coupled such that an effective modification of the gain in particular modes is achieved. Evidently, the NOR gate operation is enabled for co-propagating and counter-propagating beams \cite{Kumar-OptExpress-2006, Son-ElectronLett-2006}. While the XGM deals with the imaginary part of the nonlinear susceptibility, cross-phase modulation implies a change in the refractive index through free carriers density modulation modifying the dispersion of an amplifier active medium. Therefore, the other co-propagating beam undergoes spectral chirp, making negation functionality feasible at high repetition rates. Properly filtering the output signal, NOR gates have been realised allowing for signal processing as high as 40 Gb/s \cite{Dong-OptComm-2008}.

\textbf{Alternative platforms.} Optical fibers can be used to implement negation functionality and universal gate operation. According to the approach considered in \cite{Lai-OptExpress-2008}, one can process the optical signals propagating in a fiber through specific clock pulses propagating along the same fiber but having a slightly different frequency. Via the process of non-degenerate four-wave mixing (NFWM) new components of the signal beam are formed, effectively reducing the initial intensity. Following this way, an all-optical NOT gate has been successfully implemented. Another nonlinear effect utilized for the negation functionality of an output optical signal in birefringent fibers is nonlinear polarization rotation (NPR). A linearly polarized pump beam induces a nonlinear phase shift between two polarization axes of a birefringent fiber. Along with the pump beam, an additional probe beam is injected, having the polarization vector rotated by 45$^{\circ}$ with respect to the pump. When the intensity of the pump beam is adjusted properly, the probe polarization vector undergoes 90$^{\circ}$ rotation. Therefore, by placing a polarizer in the output port, one can observe NOR gate functionality, as was reported in \cite{Wang-AsiaComm-2011}. One of the major drawbacks of these approaches is the necessity to cope with dispersive pulse broadening, which significantly limits the speed of signal processing. Moreover the high optical intensities required together with the latency and the footprint make fibers ineffective for such applications. Alternatively, silicon wire waveguides with strong optical confinement can be utilized. The small effective modal area ($\textless$0.1 $\mathrm{\upmu{}m}^{2}$) reduces the intensity for nonlinear control signals. In \cite{Liang-OptComm-2006} authors harness two-photon absorption (TPA) in silicon to build an all-optical NOR gate. According to \cite{Liang-OptComm-2006} two encoded signal pulses being injected in the waveguide allow for modulation of a continuous light wave. Relying on this principle, a NOR gate with 80 Gb/s operational speed has been demonstrated. One of the commonly-recognized limiting factors of the approach is an accumulation of free carriers due to the TPA process \cite{Liang-OptComm-2006}. Injection-locking of a Fabry-Perot diode laser is another well-developed way to build all-optical negation logic. Normally, there is a dominant lasing mode. However, once signal beams are properly injected into the cavity, it may result in the collapse of the lasing from the dominant mode to another laser mode accordingly, thus reducing the major mode intensity. Such design allows for NOR/NOT gates operation with an extremely high extinction coefficient (over 40 dB). An apparent drawback of the approach relates to a long relaxation time of population inversion required for any conventional laser. It limits the speed of signal processing to a level of 10 Gb/s \cite{Uddin-IEEE-2009}. 

All the parameters of the NOT/NOR optical gates described above are summarized in Table S1.

\clearpage
	\begin{table}
		\caption{\textbf{The list of the reported NOR/NOT optical gates based on various platforms.}}
		\colorbox{white}{
			\begin{tabular}{ P{0.8cm} P{2.5cm} P{2cm} P{2.5cm} P{2cm} P{1.7cm} }
				\hline
			Ref.& \makecell{Nonlinear \\ element}& \makecell{Modulation\\ source} &\makecell{Response \\ period (ps)*} & \makecell{Extinction \\ ratio (dB)} & \makecell{Switching \\ energy **} \\
				\hline
		\color{blue} This work & \color{blue} Organic microcavity &\color{blue} Exciton-polariton condensation & \color{blue} $\sim$0.5 & \color{blue} $\sim$20 &\makecell{ \color{blue} $\sim$10 pJ \\ \color{blue} (80 fJ)***}\\		
		\cite{Kumar-OptExpress-2006} & SOA & XGM & \textgreater10 & 10 & 150 fJ \\
		\cite{Dong-OptComm-2008} & SOA & XGM/XPM & 25 & 12 & 12 fJ \\
		\cite{Lai-OptExpress-2008} & Optical fiber & FWM & $\sim$ 12.5 & 10 & - \\
		\cite{Wang-AsiaComm-2011} & Optical fiber & NPR & 100 & - & - \\
		\cite{Liang-OptComm-2006} & Si waveguide & TPA & 12.5 & 10 & $\sim$8 pJ \\
		\cite{Uddin-IEEE-2009} & Fabry-Perot diode laser & Injection-locking & 25 & 40 & 30 fJ \\
				\hline
			\end{tabular}
		}
	\end{table}
* The response period is derived from the reported or theoretical prediction (where exists) operational speed

** The value corresponds to the input optical energy of a beam modulating a gate output

*** The value corresponds to the extinction ratio of 3 dB

\clearpage


\end{document}